\documentclass{llncs}

\usepackage{graphicx,stmaryrd}

\newcommand{\pdot}{\(\cdot\)}

\usepackage[all]{xy}
\usepackage{array}
\usepackage{amsfonts,amsmath,amssymb,alltt}
\usepackage{latexsym}

\newcommand{\fofstr}[1]{\(\llbracket\)#1\(\rrbracket\)}
\newcommand{\stroff}[1]{\(\langle\)#1\(\rangle\)}

\author{Yves Bertot\inst{1} \and  Ekaterina Komendantskaya\inst{2}}

\institute{INRIA Sophia Antipolis, France\\ \email{Yves.Bertot@inria.fr} \and
School of Computer Science, University of St Andrews, UK\\ \email{ek@cs.st-andrews.ac.uk} 
\thanks{Work is partially supported by the INRIA CORDI post-doctoral program,
the ANR project ``A3Pat'' ANR-05-BLAN-0146 and by EPSRC postdoctoral grant EP/F044046/1.}}

\title{Using Structural Recursion for Corecursion}

\newcounter{private}
\begin{document}
\maketitle

\begin{abstract}
We propose a (limited) solution to the problem of constructing 
stream values defined by recursive
equations that do not respect
the guardedness condition. 
The guardedness condition is imposed on definitions of corecursive functions in Coq,
AGDA, and other higher-order proof assistants.
In this paper, we concentrate
in particular on those non-guarded equations where recursive calls appear under functions.
We use a correspondence between streams and functions over
natural numbers to show that some classes of non-guarded definitions
can be modelled through the encoding as structural recursive functions.
In practice, this work extends the class of stream values that can be
defined in a constructive type theory-based theorem prover with inductive and coinductive
types, structural recursion and guarded corecursion.\\
\keywordname{ Constructive Type Theory, Structural Recursion, Coinductive
  types, Guarded
Corecursion, Coq}
\end{abstract}

\section{Introduction}\label{sec:intro}
Interactive theorem provers with inductive types
\cite{Agda,Coq,NPW02,HOL} provide a restricted
programming language together with a formal meta-theory for reasoning about the language.  This
language is very close to functional programming languages, so that
the verification of a program in a conventional functional programming
language can often be viewed as a simple matter of adapting the program's
formulation to a theorem prover's syntax, thus obtaining a faithful
prover-level model.  Then one can reason about this model in the theorem
prover.  This approach has inspired studies of a large collection
of algorithms, starting from simple examples like sorting algorithms
 to
more complex algorithms, like the ones used in the computation of
Gr\"{o}bner bases, the verification of the four-colour theorem, or
compilers. 

However, the prover's programming language is restricted, especially
concerning recursion.   For instance, 
\emph{structural} restriction ensures
that all programs terminate, so that values are never undefined; we give details in Section \ref{sec:termprod}.  
Approaches
to cope with potentially non-terminating programs are available, especially
by encoding domain theory as in HOLCF \cite{Reg95}, but these approaches tend to make
the description of programs more cumbersome, because the exceptional case
where a computation may not terminate needs to be covered at every stage.
An alternative is to manage a larger class of terminating functions, mainly using
well-founded recursion \cite{Nordstrom88,Slind96}, and this approach is now widely spread among all
interactive theorem provers.

A few theorem provers \cite{Agda,Coq,NPW02} also support coinduction.  Coinductive datatypes
provide a way to look at infinite data objects.
In particular, streams of data can be viewed as infinite lists.  Coinductive
datatypes also provide room for a new class of recursive objects, known
as corecursive objects.  

Termination is not required anymore for these functions, but termination
still plays a role, since every finite value should still be computable
in finite time, even if the computation involves an interaction with a
corecursive value.  This constraint boils down to a concept of
{\em productivity}.  Roughly speaking, infinite sequences of recursive calls
where no data is being produced must be avoided.  For recursive programs,
productivity is undecidable for the same reason that termination is.  For
this reason, a more restrictive criterion is used to describe corecursive
functions that are legitimate in theorem provers.

A theorem prover like Coq provides two kinds of
recursion: terminating recursion, initially based on structural recursion
for inductive types, which can also handle well-founded recursion; and productive
corecursion, based on ``guarded'' corecursion 
\cite{Coq94,Gim96}.  
Efforts have been made to extend the basic guarded corecursion in
the same spirit that well-founded recursion extends the basic structural
recursion.  We can mention \cite{GM02} and
\cite{Bert05,BK08}, which basically incorporate 
well-founded recursion to make sure several non-productive
recursive calls are allowed as long as they ultimately become productive.
In particular, \cite{GM02} introduces a generalization of the concept
of well-founded relation that uses an extra dimension to cover at the
same time recursive or co-recursive functions; since there is an extra
dimension, two notions of limits can be used and recursive values
can mix terminating recursive and productive co-recursive aspects
in a seamless fashion.

One essential characteristic of well-founded induction and the
complete ordered families of equivalences in \cite{GM02} is that the
well-founded relation or families of equivalences must be given
as extra data to make it possible to start the definition process.
In the alternative approach described in this paper, we want to avoid
this extra burden imposed on the user, and we attempt to develop
a methodology that remains syntactic in nature.

We will concentrate on a class of recursive definitions where
{\em mapping} functions interfere in the recursive equation, thus
preventing the recursive equation to be recognised as {\em guarded
by constructors}.
The infinite sequences of Fibonacci numbers (considered e.g., in \cite{Abel06phd}) and of natural numbers (see Example \ref{ex:nats}) are 
famous representatives of the class. Many of the corecursive 
values studied, for example, in \cite{Sij89,EWD792,EGHIK07}
fail to satisfy the guardedness 
condition, precisely because functions like {\tt map} interfere
in the recursive definition.  A very elegant method
of lazy differentiation \cite{Kar01}
also gives rise to a function of multiplication for infinite streams
of derivatives in the same class of definitions.

A simple example is the following recursive equation
(studied later as Example \ref{ex:nats}):
\begin{alltt}
nats = 1::map S nats
\end{alltt}
A quick analysis shows that we can use this equation to infer the value of each
element in the stream: the first value is given directly, 
the second element is obtained from the first one
through the behaviour of the {\tt map} function, and so on.  This recursive
equation is a legitimate specification of a stream, and it can actually
be used as a definition in a conventional lazy functional programming language
like Haskell.

Thus the question studied in this article is: given a recursive equation
like the one concerning {\tt nats}, can we build a corecursive value
that satisfies this equation, using only structural recursion and guarded
corecursion?  We will describe a partial solution to this problem.  We
will also show that this solution can incorporate other interfering functions
than {\tt map}.  In Section \ref{sec:guard}, we briefly overview the class of the functions we target. 

Our proposed approach is to map every stream value to a
function over natural numbers in a reversible way: a stream {
\(s_0\)::\(s_1\)::\(\cdots\)} is mapped to the function
\fofstr{\(s\)} : \(i \mapsto s_i\), and the reverse map is an easily
defined guarded corecursive function.  It appears
that all legitimate guarded corecursive values are
mapped to structurally recursive functions and that the question of
productivity is transformed into a question of termination.
We discuss it in Section \ref{sec:rec}.

Moreover, uses of the
{\tt map} function and similar operations 
are transformed into program fragments that still respect the
constraints of structural recursion.  Thus, there are stream values
whose recursive definitions as streams are mapped to structural
recursive definitions, even though the initial equations did not respect
guardedness constraints.  For these stream values, we propose
to define the corresponding recursive function using structural
recursion, and then to produce the stream value using the reverse map
from functions over natural numbers to streams. We present this method
in Section \ref{sec:corec}.

\section{Structurally Recursive Functions}\label{sec:termprod}     

We start with defining the notions of inductive and coinductive types, and recursive/corecursive functions.
We will use the syntax of Coq throughout.  For a more detailed introduction to Coq, see \cite{BC04}. 
One can also handle inductive and coinductive types within HOL (proof assistant Isabelle) \cite{Paul97}, and within Martin-L\"{o}f
type theory (proof assistant AGDA) \cite{Agda}. 

Inductive data types are defined by introducing 
a few basic constructors that generate elements of the new type.
\begin{definition}\label{ex:nat}
The definition of the inductive type of  natural numbers is built using two constructors \texttt{O} and \texttt{S}:
\begin{verbatim}
Inductive nat : Set := O : nat | S : nat -> nat.
\end{verbatim} 
\end{definition}
This definition also implies that the type supports both pattern-matching
and recursion: on the one hand, all values in the type are either of the form
{\tt O} or of the form {\tt (S x)}; on the other hand, all values are finite
and a function is well defined when its value on \texttt{O} is given and
the value for {\tt S x} can be computed from the value for \texttt{x}.

After the inductive type is defined, one can define its inhabitants
and functions on it.  Most functions defined on the inductive type must
be defined recursively, that is, by describing values for
different patterns of the constructors and by allowing calls to the
same function on variables taken from the patterns.  

\begin{example}
The recursive function below computes the n-th Fibonacci number.
\begin{verbatim}
Fixpoint fib (n:nat) : nat :=
  match n with
  | O => 1
  | S O => 1
  | S (S p as q) => fib p + fib q
  end.
\end{verbatim}
\end{example}

There is one important property we wish every function defined in Coq
to possess: it is termination. To guarantee this, Coq uses a syntactic restriction
on definitions of functions, called \emph{structural recursion}.  
A \emph{structurally recursive} definition is such that every recursive call
is performed on a structurally smaller argument. 
The function \texttt{fib} is {\em structurally recursive}: all recursive calls are made
on variables (here {\tt p} and {\tt q})
that were obtained through pattern-matching from the initial
argument.

There are many useful functions and algorithms that are not structurally recursive, but general recursive.
They are not accepted by Coq or similar proof assistants directly, but
they can be defined using various forms of well-founded induction or induction
with respect to a {\em predicate} \cite{BC04,Bove02phd}.

It is perhaps worth mentioning that there exists an approach to termination
called ``type-based termination'' \cite{Abel06phd,BFGPU04,HPS96}. The essence
of
different methods proposed under this name is rejection of the structural
recursion as being a too restrictive and narrow method for guaranteeing
termination. Instead, this job is delegated to sized higher-order types. 
The type-based
termination promises to be a powerful tool, but it is not easy to 
implement it. As for today, the major proof assistants still rely on
structural recursion. Some non-guarded functions we formalise in this paper,
can also be handled by methods of type-based termination. However, yet it 
gives
little from the point of view of practical programming and automated proving.
Therefore the value of this paper, as well as (e.g.) \cite{BC04,BK08,Bove02phd}
is in the technical elegance and 
practical implementation in the existing proof assistants.

\section{Guardedness}\label{sec:guard}
We now consider corecursion.


The following is the definition of a coinductive type of infinite streams,
built using one constructor \texttt{Cons}.
\begin{definition}\label{ex:str}
The type of streams is given by
\begin{verbatim}
CoInductive Stream (A:Set) : Set :=
   Cons: A -> Stream A -> Stream A.
\end{verbatim}
\end{definition}
In the rest of this paper, we will write \texttt{a::tl} for 
\texttt{Cons \_ a tl}, leaving the argument {\tt A} to be
inferred from the context.

While a structurally recursive function is supposed to rely on an inductive
type for its domain and is restricted in the way recursive calls are using
this input, a corecursive function is supposed to rely on a co-inductive type
for its co-domain and is restricted in the way recursive calls are used
for producing the output.

\begin{definition}[Guardedness]
A position in an expression
is {\em pre-guarded} if it occurs as the root of the expression,
or if it is a direct
sub-term of a pattern-matching construct or a conditional statement, 
which is itself in
a pre-guarded position.  

A
position is guarded
if it occurs as a direct sub-term of a constructor for the co-inductive 
type that is being defined
and if this constructor occurs in a pre-guarded position or a guarded 
position.  A corecursive function is
\emph{guarded} if all its corecursive calls occur in guarded 
positions.
\end{definition}

\begin{example}\label{ex:map}
The coinductive function \texttt{map} applies a given function
\texttt{f} to a given infinite stream.
\begin{verbatim}
CoFixpoint map (A B :Type)(f: A -> B)(s: Stream A): Stream B :=
match s with x::s' => f x::map A B f s' end.
\end{verbatim}
In this definition's right-hand side
the match construct and the
expression {\tt f x::...} are in pre-guarded positions, the expression
{\tt map A B f s'} is in guarded position, and the definition is guarded.
\end{example}
\begin{example}\label{ex:nums}
The coinductive function \texttt{nums} takes as argument  a natural number $n$  
and produces a stream of natural numbers starting from $n$. 
\begin{verbatim}
CoFixpoint nums (n: nat): Stream nat := n::nums (S n).
\end{verbatim}  
In this definition's right-hand side, the expression {\tt n::nums (S n)}
is in a pre-guarded position, the expression {\tt nums (S n)} is in a
guarded position.
\end{example}
\begin{example}\label{ex:Zip}
The following function \texttt{zipWith} is guarded:
\begin{verbatim}
CoFixpoint zipWith (A B C: Set)(f: A -> B -> C)
     (s: Stream A)(t: Stream B) : Stream C :=
match (s, t) with  (x :: s',  y :: t')  => 
(f x y):: (zipWith A B C f s' t')
end. 
\end{verbatim}
\end{example}

Informally speaking, the guardedness condition insures that 
\begin{itemize}
\item[*]\label{it:a} each corecursive call is made under at least one constructor;
\item[**]\label{it:b} if the recursive call is under a constructor, it does not appear as an argument of any function.
\end{itemize}

Violation of any of these two conditions makes a function non-guarded. According to the two guardedness conditions above, 
we will be talking about the two classes of non-guarded functions
 - (*) and (**).

A more subtle analysis of the corecursive functions that fail to satisfy the guardedness condition * can be found in \cite{GM02,Bert05,Niq06,BK08}.
In particular, the mentioned papers offer a solution to the problem of formalising productive corecursive functions of this kind.

Till the rest of the paper, we shall restrict our attention to the second
class of functions.
To the extent of our knowledge, this paper is the first attempt to
systematically 
formulate
 the functions of this class in the language of a higher-order proof assistant with guarded
corecursion.   
\begin{example}\label{ex:nats}
Consider the following equation:
\begin{alltt}
nats = 1::map S nats
\end{alltt}
This definition is not guarded, the expression {\tt map S nats} occurs in
a guarded position, but {\tt nats} is not; see
the guardedness condition **.
Despite of this, the value \texttt{nats} is well-defined.
\end{example}

\begin{example}\label{ex:fib}
The following definition describes the stream of Fibonacci numbers:
\begin{verbatim}
fib =  0 ::  1 :: (zipWith nat nat plus (tl fib) fib).
\end{verbatim}
Again, this recursive equation fails to satisfy **.
\end{example}

\begin{example}\label{ex:dtimes}
The next example shows the function \texttt{dTimes} that multiplies the sequences of
derivatives in the elegant method of lazy differentiation of \cite{Kar01,Br08}.
\begin{verbatim} 
dTimes x y =  match x, y with 
  | x0 :: x', y0 :: y' =>
  (x0 * y0) :: (zipWith Z Z plus (dTimes x'  y)  (dTimes x y'))
   end.
\end{verbatim}
Again, this function fails to satisfy **.
\end{example}
In the next section, we will develop a method that makes it
possible to express Examples \ref{ex:nats} - \ref{ex:dtimes} as guarded corecursive values.

Values in co-inductive types usually cannot be observed as a whole, 
because of their infiniteness.
To prove some properties of infinite streams, we use a method of observation. 
For example, to prove that the two streams are \emph{bisimilar}, 
we must observe that their first elements are the same, and continue the process with the rest.

\begin{definition}\label{ex:EqSt}
Bisimilarity is expressed in the definition of the following coinductive type:

\begin{verbatim}
CoInductive EqSt:  Stream A -> Stream A -> Prop :=
| eqst : forall (a : A) (s s' : Stream A), EqSt s s' -> 
                    EqSt (a::s)(a::s').
\end{verbatim}
\end{definition}
In the rest of this paper, we will write \texttt{\(a\)==\(b\)} for
\texttt {EqSt \(a\) \(b\)}.
The definition of \texttt{\(a\)==\(b\)} corresponds to the conventional 
notion of bisimilarity as given, e.g. in \cite{JR97}.
Lemmas and theorems analogous to the \emph{coinductive proof principle} 
of \cite{JR97} are proved in Coq and can be found in \cite{BC04}.

Bisimilarity expresses that two streams are observationally equal.  Very
often, we will only be able to prove this form of equality, but for most
purposes this will be sufficient.

\section{Soundness of recursive transformations for streams}\label{sec:rec}
In this section, we show that streams can be replaced by functions.
Because there is a wide variety of techniques to define functions, this
will make it possible to increase the class of streams we can reason
about.  Our approach will be to start with a (possibly non-guarded) recursive 
equation known
to describe a stream, to transform it systematically
into a recursive equation for a structurally recursive function,
and then to transform this function back into a stream using a
guarded corecursive scheme.

As a first step, we observe how to construct a stream from a function
over natural numbers:
\begin{definition}\label{df:stroff}  
Given a function \(f\) over natural numbers, it can be transformed
into a stream using the following function:
\begin{verbatim}
Cofixpoint stroff (A:Type)(f:nat->Type) : Stream A :=
  f 0 :: stroff A (fun x => f (1+x)).
\end{verbatim}
\end{definition}
This definition is guarded by constructors.  In the rest of this
paper, we will write \stroff{\(s\)} for \texttt{stroff \_ \(s\)} leaving
the argument {\tt A} to be inferred from the context.

The function \texttt{stroff} has a natural inverse, the function
\texttt{nth} which returns the element of a stream at a given rank:
\begin{definition}\label{df:nth} The function {\tt nth}\footnote{In Coq's library,
this function is defined under the name \texttt{Str\_nth}.}
is defined as follows:
\begin{verbatim}
Fixpoint nth (A:Type) (n:nat) (s: Stream A) {struct n}: A :=
match s with  a :: s' =>
  match n with | O =>  a | S p => nth A p s' end
end.
\end{verbatim}
\end{definition}
In the rest of this paper, we will omit the first argument (the
type argument) of {\tt nth}, following Coq's approach to implicit arguments.
We will use notation \fofstr{s} when talking about \texttt{(fun n => nth n s)}.

It is easy to prove that \fofstr{\(\cdot\)} and \stroff{\(\cdot\)} are inverse
of each other. Composing these two functions is the essence of the method we
develop here. The lemmas below are essential for guaranteeing the soundness of our method.
\begin{lemma}\label{nthstroff}
For any function \texttt{f} over natural numbers, 
\texttt{\(\forall\)n, nth n {\stroff{f}} = f n}.
\end{lemma}

\begin{lemma}\label{th:sound}
For any stream \texttt{s}, \texttt{s == \stroff{\fofstr{\(s\)}}}. 
\end{lemma}
\begin{proof}
Both proofs are done in Coq and available in \cite{BKmodcorec08}. 
\end{proof}

We now want to describe a transformation for (non-guarded) recursive equations for streams. 
A recursive equation for a stream would normally have the form
\begin{eqnarray}
a &=& e \label{eqn1}
\end{eqnarray}
where both $a$ and $e$ are streams, and $a$ can also occur in the expression $e$; see Examples \ref{ex:nats} - \ref{ex:dtimes}.
We use this initial non-guarded equation to formulate a guarded equation for $a$ of the form:
\begin{eqnarray}
 a &= &\texttt{\stroff{$e'$}} \label{eqn2}
\end{eqnarray}
where $e'$ is a function extensionally equivalent to \fofstr{$e$}.  
As we show later in this section, we often
need to evaluate \texttt{nth} only partially or only at a certain depth, 
this is why the job cannot be fully delegated
to \texttt{nth}.

The definition of $e'$ will have the form
\begin{eqnarray}
e'~n &=& E \label{eqn3}
\end{eqnarray}
where $e'$ can again occur in the expression $E$.

\begin{example}[zeroes]
For simple examples, we can go through steps (\ref{eqn1})-(\ref{eqn3})
intuitively.
Consider the corecursive guarded definition of a stream \texttt{zeroes}
that contains an infinite repetition of \texttt{0}.
\begin{verbatim}
CoFixpoint zeroes := 0 :: zeroes.
\end{verbatim}
We can model
the body of this corecursive definition as follows:

\begin{verbatim}
Fixpoint nzeroes (n:nat) : nat :=
  match n with 0 => 0 | S p => nzeroes p end.
\end{verbatim}

This is a legitimate structurally recursive
definition for a function that maps any natural number to zero. 
Note that the obtained function is extensionally 
equal to \texttt{\fofstr{zeroes}}. 

\begin{verbatim}
Lemma nth_zeroes: forall n, nth n zeroes = nzeroes n.
\end{verbatim}

Thus, a stream that is bisimilar to \texttt{zeroes} can be obtained
by the following commands:

\begin{verbatim}
Definition zeroes' := stroff _ nzeroes.
\end{verbatim}
By Lemma \texttt{nth\_zeroes} and Lemma \ref{th:sound}, \texttt{zeroes} and 
\texttt{zeroes'} are bismilar, see \cite{BKmodcorec08} for a proof.
\end{example}

The main issue is to describe a systematic transformation from the expression $e$ in the equation~\ref{eqn1} to the expression $E$ in the equation (\ref{eqn3}). This "recursive" part of the 
work will be the main focus of the next section.

\section{Recursive Analysis of Corecursive Functions}\label{sec:rec}
We continue to systematise the steps (\ref{eqn1})-(\ref{eqn3}) of the transformation for a recursive equation 
$a = e$.

The expression \(e\) can be seen as the application of a function \(F\)
to \(a\).  In this sense, the recursive definition of \(a\) expresses
that \(a\) is fixpoint of \(F\).  The type of \(F\) is {\tt Stream
  \(A\) \(\rightarrow\) Stream \(A\)} for some type \(A\).  We will derive a new
function \(F'\) of type \hbox{\tt (nat \(\rightarrow\) \(A\)) \(\rightarrow\) (nat \(\rightarrow\) \(A\))};
 the recursive function \(a'\) that we want to define is a fixed
point of \(F'\).  We obtain \(F'\) from \(F\) in two stages:

\textbf{Step 1.} We compose \(F\) on the left with \stroff{\pdot} and on
the right with \fofstr{\pdot}.  This naturally yields a new function of the
required type.  In practice, we do not use an explicit composition function,
but perform the syntactic replacement of the formal parameter with the
\stroff{\pdot} expression everywhere.

\begin{example}
For instance, when considering the {\tt zeroes} example, the initial function
\begin{verbatim}
Definition zeroes_F (zeroes:Stream nat) := 0::zeroes
\end{verbatim}
is recursively  transformed into the function:
\begin{verbatim}
Definition zeroes_F' (nzeroes : nat -> nat) :=
  nth n (0::stroff nzeroes).
\end{verbatim}
\end{example}
The corecursive value we consider may be a function
taking arguments
in types \(t_1,\ldots, t_n\), that is, the function \(F\) may actually be defined as a
function of type \hbox{\((t_1 \rightarrow \cdots \rightarrow t_n \rightarrow
  {\tt Stream} A) \rightarrow (t_1 \rightarrow \cdots \rightarrow t_n
  \rightarrow {\tt Stream} A)\)}.  The reformulated function \(F'\) that is obtained after composition with
\stroff{\pdot} and \fofstr{\pdot} has the corresponding type where
{\tt Stream \(A\)} is replaced with \({\tt nat} \rightarrow A\). Thus, it is
the first argument that incurs a type modification.
When one of the types \(t_i\) is itself a stream type, we can choose to
leave it as a stream type, or we can choose to replace it also with a
function type.  When replacing \(t_i\) with a function type,
we have to add compositions with \fofstr{\pdot} and \stroff{\pdot} at all
positions where the first argument \(f\) of \(F\) is used, to express
that the argument of \(f\) at the rank \(i\) must be converted from a stream
type to a function type and at all positions where the argument of the 
rank \(i+1\)
of \(F\) is used, to express that this argument must be converted from a
function type to a stream type.

We choose to perform this transformation of a stream argument into a
function argument only when the function being defined is later used for another
recursive definition.  In this paper, this happens only for the functions
{\tt map} and {\tt zipWith}.

\begin{example}\label{map}
Consider the function \texttt{map} from Example~\ref{ex:map}.
The function \(F\) for this case has the following form:
\begin{verbatim}
Definition map_F 
 (map : forall (A B:Type)(f: A -> B), Stream A -> Stream B) :=
 fun A B f s => match s with a::s' => f a::map A B f s' end.
\end{verbatim} 
The fourth argument to {\tt map} and the fifth argument to {\tt map\_F}
have type {\tt Stream A} and we choose to replace this type with a
function type.  We obtain the following new function:
\begin{alltt}
Definition map_F'
  (map : forall (A B:Type)(f:A -> B), (nat -> A) -> nat -> B :=
fun A B f s => \fofstr{match \stroff{s} with a::s' => f a::\stroff{map A B f \fofstr{s'}} end}.
\end{alltt}
\end{example}

\textbf{Step 2.} We go on transforming the body of \(F'\) according to
rewriting rules that express the interaction between \stroff{\pdot},
\fofstr{\pdot}, and the usual functions and constructs that deal with
streams.

The Table~\ref{tb:rules} gives a summary of the rewriting rules for
the transformation.

\begin{table}
\begin{tabular}{|p{12cm}|}
\hline
\begin{enumerate}
\item {\tt nth} \(n\) \stroff{\(f\)} = \(f~n\),
\item {\tt nth} \(n\) {\tt (\(a\)::\(s'\))} = {\tt match} \(n\) {\tt with} 0 {\tt=>} \(a\) {\tt|} {\tt S} \(p\) {\tt=> nth} \(p\) \(s'\) {\tt end},
\item {\tt hd} \stroff{\(f\)} = \(f~0\),
\item {\tt tl} \stroff{\(f\)} = \stroff{{\tt fun \(n\) => \(f\) (S \(n\))}},
\item {\tt match \stroff{\(f\)} with \(a\)::\(s\) => \(e~a~s\) end} =
{\tt \(e~(f~0)\)~\stroff{\tt fun \(n\) => \(f\) (S \(n\))}},
\item \(\beta\)-reduction.
\end{enumerate}
\setcounter{private}{\value{enumi}}\\
\hline
\end{tabular}
\caption{Transformation rules for function representations of streams.}\label{tb:rules}
\end{table}
All these rules can be proved as theorems in the theorem prover
\cite{BKmodcorec08}: this
guarantees soundness of our approach.  However, this
kind of rewriting cannot be done directly inside the theorem prover, since
rewriting can only be done while proving statements, while we are in the
process of defining a function.  Moreover, the rewriting operations must be
done thoroughly, even inside lambda-abstraction, even though an operation for that may
not be supported by the theorem prover (for instance, in the calculus of
constructions as it is implemented in Coq, rewriting does not occur inside
abstractions).

The rewriting rules make the second argument of {\tt nth}
decrease.  When the recursive stream definition is guarded, this process
ends with a structural function definition.

\begin{example} Let us continue with the definition for {\tt map}.
\begin{alltt}
map_F' map A B f s n =
  nth n match \stroff{s} with a:: s' => f a::\stroff{map A B f \fofstr{s'}}end
= nth n (f(s 0)::\stroff{map A B f \fofstr{s'}}) end
= match n with
    0 => f(s 0) | S p => nth p \stroff{map A B f \fofstr{\stroff{fun n => s (S n)}}}
  end
= match n with 0\,=>\,f(s 0)|\,S p\,=>\,map A B f \fofstr{\stroff{fun n => s (S n)}}\,p\,end
= match n with 0\,=>\,f(s 0)|\,S p\,=>\,map A B f (fun n => s (S n)) p\,end
\end{alltt}
\end{example}
When considered as the body for a recursive definition of a function {\tt map'},
the last right-hand side is a good structural recursive definition with respect
to the initial parameter {\tt n}.  We can use this for a structural definition:
\begin{verbatim}
Fixpoint map' (A B : Type) (f : A -> B) (s: nat -> A)
     (n : nat) {struct n} :=
 match n with 0 => f a | S p => map' A B f (fun n => s (S n)) p end.
\end{verbatim}
This function models the {\tt map} function on streams, as a function on
functions.  It enjoys a particular property, which plays a central role in
this paper:
\begin{lemma}[Form-shifting lemmas]\label{lem:rmapr}
\begin{alltt}
\(\forall\) \(f\) \(s\) \(n\), nth \(n\) (map \(f\) \(s\)) = \(f\) (nth \(n\) \(s\))

\(\forall\) \(f\) \(s\) \(n\), \fofstr{map} \(f\) \(s\) \(n\) = \(f\) (\(s\) \(n\)).
\end{alltt}
\end{lemma}
\begin{proof} See \cite{BKmodcorec08}.
\end{proof}
Thanks to the second statement of the lemma, \(s\) can be moved from an
argument position to an active function position, as will later be needed
for verifying structural recursion of other values relying on \texttt{map}.

Now, we show the same formalisation for the function \texttt{zipWith}:
\begin{example}[Zip] 
The function \texttt{zipWith} can also be transformed, with the choice that
both stream arguments are transformed into functions over natural
numbers.
\begin{verbatim}
Definition zipWith_F
  (zipWith : forall (A B C : Type), (A -> B -> C) ->
                               Stream A -> Stream B -> Stream C)
  (A B C : Type)(f : A -> B -> C)(a : Stream A)(b : Stream B) :=
  match a, b with 
    x :: a', y :: b' => f x y :: zipWith A B C f a' b'
  end.
\end{verbatim}
Viewing arguments {\tt a} and {\tt b} as functions and applying the rules
from Table \ref{tb:rules} to this definition yields the following recursive
equation:
\begin{alltt}
zipWith_F' zipwith' A B C f a b n =
  match n with
    0 => f (a 0) (b 0)
  | S p => zipwith' (fun n => a (S n)) (fun n => b (S n)) p
  end
\end{alltt}
\end{example}

Here again, this leads to a legitimate structural recursive definition on the
fourth argument of type {\tt nat}.  We also have form-shifting lemmas:
\begin{lemma}[Form-shifting lemmas]
\begin{alltt}

\(\forall\) \(f\) \(s\sb{1}\) \(s\sb{2}\) \(n\), \texttt{nth} \(n\) (\texttt{zipWith} \(f\) \(s\sb{1}\) \(s\sb{2}\)) = f (\texttt{nth} \(n\) \(s\sb{1}\)) (\texttt{nth} \(n\) \(s\sb{2}\))

\(\forall\) \(f\) \(s\sb{1}\) \(s\sb{2}\) \(n\), \fofstr{\texttt{zipWith}} \(f\) \(s\sb{1}\) \(s\sb{2}\) \(n\) = \(f\) (\(s\sb{1}\) \(n\)) (\(s\sb{2}\) \(n\)).
\end{alltt}
\end{lemma}
\begin{proof}
See \cite{BKmodcorec08}.
\end{proof}  
The second statement also moves \(s_1\) and \(s_2\) from argument position to
function position. 

Unfortunately, we do not know the way to automatically discover the
Form-shifting lemmas; although the statements of these lemmas follow the same
generic pattern and once stated, the proofs for them do not tend to
be difficult.
Instead, as we illustrate in the Conclusion, we sometimes can give a convincing
argument showing that a form-shifting lemma for a particular function cannot
be found; and this provides an evidence that our method is not applicable to
this function.
That is, existence or non/existence of the form-shifting lemmas 
can serve as a criterium for determining whether the function can be covered
by the method.

\section{Satisfying Non-Guarded Recursive Equations}\label{sec:corec}
Form-shifting lemmas play a role when studying recursive equations that do not satisfy the guardedness condition **, that is, when
the corecursive call is made under functions like \texttt{map} or \texttt{zipWith}.
To handle these functions, we simply need
to add one new rule, as in Table~\ref{tb:rules2}, which
will handle occurrences of each function that has a form-shifting lemma.
\begin{table}
\begin{tabular}{|p{12cm}|}
\hline
\begin{enumerate}
\setcounter{enumi}{\value{private}}
\item Let \texttt{f} be a function and \texttt{C} be a context in which
arguments of \texttt{F} appear. 
If a form shifting lemma has the following shape:
\[\forall a_1~\cdots~a_k~s_1~\cdots~s_l~n,
\texttt{\fofstr{f}}
a_1~\cdots~a_k~s_1~\cdots~s_l ~n
= C[a_1,\ldots,a_k,s_1~n,\ldots,s_l~n],\]
then this equation should be used as an extra rewriting rule.
\end{enumerate}\\
\hline
\end{tabular}
\caption{Rule for recursive transformation of non-guarded streams}\label{tb:rules2}
\end{table}

The extended set of transformation rules from Tables~\ref{tb:rules}
and~\ref{tb:rules2} can now be used to produce functional definitions of streams
that were initially defined by non-guarded corecursive equations.  The technique is
as follows:
\begin{enumerate}
\item[(a)] Translate the equation's right-hand-side as prescribed by the
rules in Tables~\ref{tb:rules} and~\ref{tb:rules2},
\item[(b)] Use the equation as a recursive definition for a function,
\item[(c)] Use the function \stroff{\(\cdot\)} to obtain the corresponding
stream value,
\item[(d)] Prove that this stream satisfies the initial recursive equation,
using bisimilarity as the equality relation.
\end{enumerate}
For the last step concerning the proof, we rely on two features provided
in the Coq setting:
\begin{itemize}
\item For each recursive definition, the Coq system can generate
a specialised induction principle, as described in \cite{BartheCourtieu02},
\item A proof that two streams are bisimilar can be transformed into
a proof that their functional views are extensionally equal, using
the theorem \texttt{ntheq\_eqst}:
\begin{alltt}
ntheq_eqst :
  \(\forall\)A (s1 s2:Stream A), (\(\forall\)n, nth n s1 = nth n s2) -> s1 == s2
\end{alltt}
\end{itemize}
Using these two theorems and combining them with systematic rewriting
with Lemma~\ref{nthstroff} and the form-shifting lemmas, we
actually obtain a tactic we called
\texttt{str\_eqn\_tac} in \cite{BKmodcorec08} which proves the
recursive equations automatically.

We illustrate this method using our running examples.
\begin{example}
Consider the corecursive non-guarded definition of \texttt{nats}
from Example \ref{ex:nats}.  Here is the initial equation
\begin{alltt}
 nats = 1::map S nats
\end{alltt}
After applying all transformation rules we obtain the following equation
between functions:
\begin{alltt}
\fofstr{nats} = fun n => if n = 0 then 1 else S (\fofstr{nats} (n - 1)).
\end{alltt}  
This is now a legitimate structurally recursive equation defining
\texttt{\fofstr{nats}}, from which we define \texttt{nats} as
\texttt{nats = \stroff{\fofstr{nats}}}.
The next step is to show that \texttt{nats} satisfies the equation of
Example~\ref{ex:nats}.
\begin{alltt}
nats == 1::map S nats
\end{alltt}
Using the theorem \texttt{ntheq\_eqst} and Lemma \ref{nthstroff} on 
the left-hand-side this reduces
to the following statement:
\begin{alltt}
\(\forall\) \(n\), \fofstr{nats} \(n\) = nth \(n\) (1::map S \stroff{\fofstr{nats}})
\end{alltt}
We can now prove this statement by induction on the structure of the
function \texttt{\fofstr{nats}}, as explained in \cite{BartheCourtieu02}.
This gives two cases:
\begin{alltt}
1 = nth 0 (1::map S \fofstr{nats})

S (\fofstr{nats} p) = nth (S p) (1 :: map S \stroff{\fofstr{nats}})
\end{alltt}
The first goal is a direct consequence of the definition of
{\tt nth}.  The second goal reduces as follows:
\begin{alltt}
S (\fofstr{nats} p) = nth p (map S \stroff{\fofstr{nats}})
\end{alltt}
Rewriting with the first form-shifting lemma for {\tt map} yields
the following goal:
\begin{alltt}
S (\fofstr{nats} p) = S (nth p \stroff{\fofstr{nats}})
\end{alltt}
Rewriting again with Lemma~\ref{nthstroff} yields the following trivial
equality.
\begin{alltt}
S (\fofstr{nats} p) = S (\fofstr{nats} p).
\end{alltt}
\end{example}

\begin{example}
The sequence of Fibonacci numbers can be defined by the following equation:
\begin{alltt}
fib = 1::1::zipWith plus fib (tl fib)
\end{alltt}
When processing the left-hand side of this equation using the rules from Tables \ref{tb:rules},
\ref{tb:rules2} and the form-shifting lemma for {\tt zipWith}, we obtain the following
code:
\begin{alltt}
\fofstr{fib} = fun n =>
 match n with
 | 0 => 1
 | S p => match p with 0\,=>\,1\,|\,S q\,=>\,\fofstr{fib} q + \fofstr{fib} (1+q) end
 end
\end{alltt}
This is still not accepted by the Coq system
because \texttt{(1+q)} is not a variable term, however it is semantically equivalent to \texttt{p}, and the following text is accepted:
\begin{alltt}
\fofstr{fib} = fun n =>
 match n with
 | 0 => 1
 | S p => match p with 0\,=>\,1\,|\,S q\,=>\,\fofstr{fib} q + \fofstr{fib} p end
 end
\end{alltt}
Again, by Definition \ref{df:stroff}, we can define a stream
\texttt{fib = \stroff{\fofstr{fib}}}, and \texttt{fib} is proved to satisfy
the initial
recursive equation automatically.
\end{example}
It is satisfactory that we have a systematic method to produce
a stream value for the defining recursive equation, but we should
be aware that the implementation of \texttt{fib} through a structural
recursive function does not respect the intended behaviour and has a much
worse complexity ---exponential--- while the initial equation can be implemented
using lazy data-structures and have linear complexity.

Finally, we illustrate the work of this method on the function \texttt{dTimes} from Example \ref{ex:dtimes}:
\begin{example}\label{ex:dtimes2}
For the function \texttt{dTimes}, we choose to leave the two stream arguments
{\tt x} and {\tt y} as streams.  We recover the structurally recursive function \texttt{\fofstr{dTimes}} from Example \ref{ex:dtimes}:
\begin{alltt}
\fofstr{dTimes} (x y:Stream nat) (n:nat)\{struct n\} = 
  match x, y with
  | x0 :: x', y0 :: y' => 
    match n with 
    | 0 => x0 * y0
    | S p => (\fofstr{dTimes} x' y p) + (\fofstr{dTimes} x y' p)
    end
  end.
\end{alltt}
It remains to define the stream \texttt{\stroff{\fofstr{dTimes}}}, which is a straightforward application of Definition \ref{df:stroff}, 
and to prove that it satisfies the initial recursive equation from Example \ref{ex:dtimes}.
In \cite{BKmodcorec08}, the proof is again handled automatically.
\end{example}
\section{Conclusions}\label{sec:concl}

The practical outcome of this work is to provide an approach to model
corecursive values that are not directly accepted by
the ``guarded-by-constructors'' criterion, without relying on more
advanced concepts like well-founded recursion of ordered families of
equivalences.  With this approach we can
address formal verification for a wider class of functional
programming languages.  The work presented here is complementary to the
work presented in \cite{BK08}, since the method in that paper only considers
definitions where recursive calls occur outside of any constructor, while
the method in this paper considers definitions where recursive calls are inside
constructors, but also inside interfering functions.

The attractive features of this approach is that it is systematic and simple.
It appears to be simpler than, e.g., related work done in \cite{GM02,BK08,Dan08}
that involved introducing particular coinductive types and manipulating
ad-hoc predicates.   Although
the current state of our experiments relies on manual operations, we believe
the approach can be automated in the near future, yielding a command
in the same spirit as the {\tt Function} command of Coq recent versions.

The Coq system also provides a mechanism known as extraction which produces
values in conventional functional programming languages.  When it comes
to producing code for the solution of one of our recursive equations on
streams, we have the choice of using the recursive equation as a definition,
or the extracted code corresponding to the structurally recursive model.
We suggest that the initial recursive equation, which was used as
our specification, should be used as the definition, because  the
structural recursive value may not respect the intended computational
complexity.  This was visible in the model we produced
for the Fibonacci sequence, which does not take advantage of the 
value re-use as described in the recursive equation.  We still need to
investigate whether using the specification instead of the code will be
sound with respect to the extracted code.

The method presented here is still very limited: it cannot cope
with the example of
the Hamming sequence, as proposed in \cite{EWD792}.  A recursive definition
of this stream is:
\begin{alltt}
  H = 1::merge (map (Zmult 2) H) (map (Zmult 3) H)
\end{alltt}
In this definition, {\tt merge} is the function that takes two streams and
produces a new stream by always taking the least element of the two
streams: when the input streams
are ordered, the output stream is an ordered enumeration of all values in
both streams.  Such a {\tt merge} function is easily defined by guarded
corecursion, but \texttt{merge} interferes in the definition of {\tt H} in the same
way that {\tt map} interfered in our previous examples. This time, we do not
have any good form-shifting lemma for this function.  The hamming sequence
can probably be defined using the techniques of \cite{GM02} and we
were also able to find another syntactic approach for this example,
this new approach is a subject for another paper.

\bibliographystyle{abbrv}
\bibliography{Coq}

\end{document}